\documentclass{iau}

\usepackage{natbib}
\usepackage{amsmath}
\usepackage{graphicx}
\usepackage{multirow}
\usepackage{listings}
\usepackage[symbol]{footmisc}

\begin{document}

\lefttitle{M. Lopez, V. Boudart, S. Schmidt and S. Caudill}
\righttitle{Simulating transient noise burst with \texttt{gengli}}

\jnlPage{1}{7}
\jnlDoiYr{2021}
\doival{10.1017/xxxxx}

\aopheadtitle{Proceedings IAU Symposium}
\editors{C. Sterken,  J. Hearnshaw \&  D. Valls-Gabaud, eds.}

\title{Simulating transient burst noise with \texttt{gengli}}

\author{Melissa Lopez$^{1,2}$, Vincent Boudart$^{3}$, Stefano Schmidt$^{1,2}$, Sarah Caudill$^{1,4,5}$}

\affiliation{$^{1}$Institute for Gravitational and  Subatomic Physics (GRASP), Utrecht University, Princetonplein 1, 3584 CC Utrecht, The Netherlands}
\affiliation{$^{2}$Nikhef, Science Park 105, 1098 XG, Amsterdam, The Netherlands}
\affiliation{$^{3}$STAR Institute, Bâtiment B5, Université de Liège, Sart Tilman B4000 Liège, Belgium}
\affiliation{$^{4}$Department of Physics, University of Massachusetts, Dartmouth, MA 02747, USA}
\affiliation{$^{5}$Center for Scientific Computing and Visualization Research, University of Massachusetts, Dartmouth, MA 02747, USA}

\begin{abstract}
In the field of gravitational-wave (GW) interferometers, the most severe limitation to the detection of transient signals from astrophysical sources comes from transient noise artefacts, known as glitches, that happens at a rate around $1$ per minute. Because glitches reduce the amount of scientific data available, there is a need for better modelling and inclusion of glitches in large-scale studies, such as stress testing the search pipelines and increasing the confidence of detection.
In this work, we employ a Generative Adversarial Network (GAN) to produce a particular class of glitches ({\it blip}) in the time domain. We share the trained network through a user-friendly open-source software package called \texttt{gengli} and provide practical examples of its usage.
\end{abstract}


\maketitle

\vspace{-5mm}
\section{Introduction}

%%%%---------------------------------------------------------------
{The existence of gravitational-wave (GW) signals  was successfully proven by LIGO and Virgo collaborations during the first observing run (O1)} \citep{FirstGW}. After an upgrade of the detectors to increase their sensitivity, Advanced LIGO \citep{LIGOScientific:2014pky} started in November $2016$ the second observing run (O2), which Advanced Virgo \citep{VIRGO:2014yos} joined in August $2017$ \citep{O2_LIGO_Virgo}. Following significant upgrades, in April $2019$, the third observing run (O3) was initiated by LIGO-Virgo collaboration \citep{O3_LIGO_Virgo, LIGOScientific:2021djp}. In the coming years, the improved second generation of interferometers and the construction of the third generation of detectors, such as Einstein Telescope \citep{Einstein_telescope, Science_case_ET}, will increase significantly the detection sensitivity.

%%%%----------------------------------------------------

\begin{figure}[!t]
\centering
\includegraphics[width=0.6\textwidth]{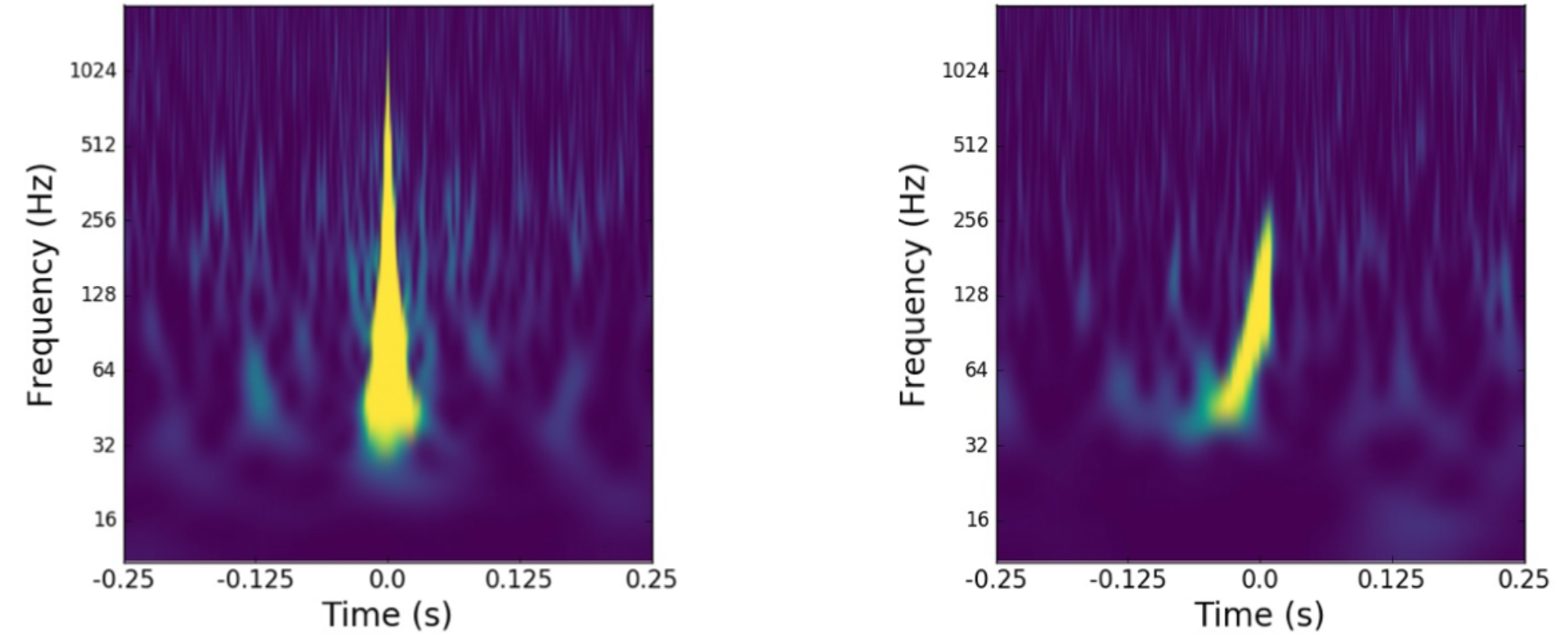}
\caption{A blip glitch \textit{(left)} that is similar to an event with total mass $106.6^{+13.5}_{-14.8} M_{\odot}$ \textit{(right)}.}
\label{fig:glitch_IMBH}
\end{figure}

Despite the significant improvements to isolate the detectors from non-cosmic disturbances, they are still susceptible to non-Gaussian noise, known as ``glitches", which come in a wide variety of time-frequency morphologies and are produced by instrumental or environmental causes. They reduce the amount of analyzable data, biasing astrophysical detection, and mimicking GW signals \citep{noise_characterization}, as in Fig. \ref{fig:glitch_IMBH}. Thus, it is fundamental to identify and characterize them for their elimination. 

%%%%----------------------------------------------------

Due to the overwhelming amount of glitches present in the LIGO data (\cite{LIGOScientific:2021djp}), identifying them goes beyond human ability.
An exciting solution is to construct machine learning (ML) algorithms to classify their different morphologies. With this idea in mind, \cite{Gravity_spy} combine the strengths of both humans and computers to analyze and characterize LIGO glitches. Through the Zooniverse platform, volunteers provide large labelled data sets to train an ML algorithm, called \textit{Gravity Spy}, while the ML algorithm learns to classify the rest of the glitches correctly and provides feedback to the participants. In practice, we feed to the ML classifier the glitch time series that we wish to classify. The Q-transform of the input is created and fed to the ML classifier, that assigns a label and a confidence value $c_{GS}$, where $c_{GS}$ is the classification probability (see \cite{Gravity_spy}).

%%%%----------------------------------------------------

While the identification of glitches is the first step towards their robust mitigation, in this investigation we intend to \textit{generate} the known classes of glitches with Generative Adversarial Networks (GAN) to further understand their principal features. This does not only allow us to enhance our understanding of their morphologies but it can also be used for various applications in GW data analysis, such as mock data challenges (see  \cite{Lopez:2022lkd} for a discussion). In this work, that accompanies the main publication \cite{Lopez:2022lkd}, we present a flexible and user-friendly tool \textit{gengli} \footnote[2]{ The code is released as a Git repository in \href{https://git.ligo.org/melissa.lopez/gengli}{melissa.lopez/gengli}} for glitch generation.

%%%%---------------------------------------------------

%%%%----------------------------------------------------

\section{Data}

To train our ML model, we select blips from Livingston (L1) and Hanford (H1) that have a confidence $c^1_{GS} \geq 0.9 $, during O2 \footnote[3]{The data is accessible through the GWOSC  (\url{ https://www.gw-openscience.org/data/})}. However, since the glitch is surrounded by non-stationary and uncorrelated noise, there is little structure that our ML method can retrieve. Therefore, we need to extract glitches from the input strain.
For this aim, we employ BayesWave \citep{BayesWave} to fit and reconstruct the input signal with data-driven wavelet models.
The {\it whitening} operation is defined on the frequency domain glitch $\tilde{g}$ as: $ \small \tilde{g}_W = {\tilde{g}}/{S_n(f)},$ where $S_n(f)$ is the power spectral density (PSD) of the input data.

%%%%----------------------------------------------------

The input provided to BayesWave is a time series of $2.0$ s at $4096$ Hz, where the blip is centred. After the reconstruction, we crop the output to a size of $938$ data points, removing irrelevant data.
We evaluate the quality of the reconstruction by injecting the output in real whitened noise, and re-evaluating with {Gravity Spy} to select blip glitches with $c^2_{GS} \geq 0.9$.

%%%%----------------------------------------------------

\begin{figure}[t]
\centering
\includegraphics[width=0.9\textwidth]{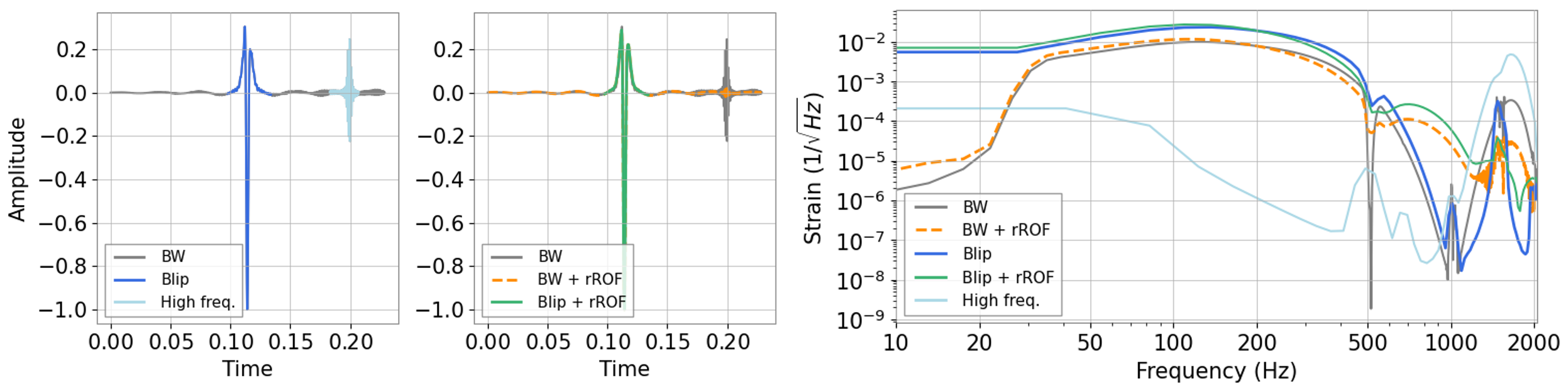}
\caption{\textit{(Left)} Reconstructed blip with BayesWave. \textit{(Middle)} Denoised reconstructed blip with rROF. \textit{(Right)}  Resulting amplitude spectral density (ASD) for reconstructed and denoised blips.}
\label{fig:ex_denoised}
\end{figure}

The reconstruction of BayesWaves is not perfect since there is still some noise contribution at high frequencies (light blue) that will hinder the learning of our ML algorithm, (see Fig. \ref{fig:ex_denoised} \textit{(left)}).  To minimize this contribution we employ regularized Rudin-Osher-Fatemi (rROF) proposed in \cite{rROF}, which solves the denoising problem as a variational method. 
In Fig. \ref{fig:ex_denoised} (middle), we plot the BayesWaves reconstruction denoised with rROF (dashed orange), and the denoised characteristic blip (green). In Fig. \ref{fig:ex_denoised} (right), we show the amplitude spectral density (ASD) of the BayesWaves reconstruction with and without denoising (grey and dashed orange), as well as the characteristic peak with and without denoising (blue and green) and the original high-frequency contribution (light blue). We can maintain the structure of the characteristic peak by damping the power of the high-frequency contribution.

\section{Methodology}

%%%%---------------------------------------------------------------

\subsection{Generative Adversarial Networks}

%%%%---------------------------------------------------------------

GANs, first introduced by \cite{Original_GAN}, are a class of generative algorithms in which two networks compete against each other to achieve realistic data generation. 
One network is responsible for the generation of new data from random noise, while the other tries to discriminate the generated samples from the real training data. The generator learns the features of real data that should be mimicked to ``foul" the discriminator, and stores them in a latent space. At the end of the training, new samples are drawn by randomly taking a latent space vector and passing it to the generator, which translates this into real data.

%%%%---------------------------------------------------------------

The work from \cite{Original_GAN} suffers from two major problems, known as \textit{vanishing gradients} and \textit{meaningless loss function} \citep{Weng2019FromGT}.
To address those issues \cite{Arjovsky} developed Wasserstein GANs, where they use the Earth's mover distance or Wasserstein-1 distance ($W_1$) as a loss function. This change led to reformulate the optimization problem as $\tiny \theta_{opt} = arg\, \underset{\theta}{min}\, W_1(P_{x}\|P_{\Tilde{x}})$,  where $W_1$ is evaluated between the true data distribution $P_{x}$ and the fake data distribution $P_{\Tilde{x}}$. Rewriting this equation yields, 
\vspace{-2mm}
\begin{equation}
    \small\theta_{opt} = arg\, \underset{\theta}{min}\, \underset{\phi:||D(x,\phi)||_L \leq 1}{max} L(\phi,\theta) \text{\quad} \,\, \text{with}\,\, \small L(\phi,\theta) = - E_{x \sim P_{x}}\big[ D(x,\phi) \big] + E_{\Tilde{x} \sim P_{\Tilde{x}}}\big[ D(\Tilde{x},\phi) \big]
    \label{optimization problem}
\end{equation}
where $D(\phi)$ and $G(\theta)$ are for the discriminator and the generator is seen as a function of their weight. $E_{x \sim P_{x}}$ indicates that the expression has been averaged over a batch of real images. Similarly, $E_{\Tilde{x} \sim P_{\Tilde{x}}}$ indicates that the expression has been average over a batch of generated images, where $\Tilde{x}=G(z,\theta)$ and $z$ is the latent space vector. The new condition in Eq. \ref{optimization problem} imposes that the discriminator must be 1-Lipschitz continuous \citep{Arjovsky}.

To fulfil this constraint, we implement the idea by \cite{Gulrajani}, which consists in adding a regularization term to the discriminator loss, known as \textit{gradient penalty} (GP): 

\begin{equation}
    \small L_{tot} = L(\phi,\theta) + \lambda \,GP(\phi) \,with \,\small GP(\phi) = E_{\hat{x} \sim P_{\Tilde{x}}} \Big[ \big( || \bigtriangledown_x\,D(\hat{x},\phi)||_2 - 1 \big)^2 \Big] ,
    \label{gradient_penalty}
\end{equation}

where $\lambda$ is a regularization parameter, $||\cdot||_2$ denotes the L2-norm and $\hat{x}$ is evaluated following: $\small\hat{x} = \Tilde{x}\,t + x\,(1-t)$, being $x$ a real sample and $\Tilde{x}$ a fake sample, with $t$ sampled uniformly $\in [0, 1]$. However, GP cannot penalize $W_1$ everywhere. In particular, at the beginning of the training, being the generated samples quite far from the true data manifold, the Lipschitz condition is not enforced until the generator becomes sufficiently good.

To overcome this obstacle, \cite{CTGAN} have proposed a second penalization term that directly penalises the points near any observed real data point $x$. Therefore, when the gradient penalty fails to enforce the Lipschitz continuity in the close vicinity of $x$, the new term will constrain the latter. To penalize the real data manifold, \cite{CTGAN} applied their new constraint to two perturbed versions of the real samples $x$. For this, they introduced dropout layers into the discriminator architecture. The features kept by the dropout layer are random, which ultimately leads to two different estimates noted $D(x')$ and $D(x'')$. The penalty, called \textit{consistency term} (CT), is then applied to these two estimates following: 

\vspace{-1mm}
\begin{equation}
    \small CT(\phi) = E_{x \sim P_{x}}\big[ max \big( 0, \, d(D(x',\phi),D(x'',\phi)) \\ + 0.1\,d(D\_(x',\phi),D\_(x'',\phi)) - M' \big) \big]
\end{equation}
where d(.,.) is the L2 metric, $D\_$ stands for the second-to-last layer output of the discriminator and $M'$ is a constant value. The final discriminator loss in \cite{CTGAN} is then:
\vspace{-3mm}
\begin{equation}
    \small L_{D} = L(\phi,\theta) + \lambda_1\,GP(\phi) + \lambda_2\,CT(\phi)
    \label{loss D}
\end{equation}
with $\lambda_2$ being the consistency parameter, that is used to tune the weight of CT for GP. While in Eq. \ref{loss D} we update the weights of the discriminator, we update the weights of the generator as in \cite{Arjovsky}, with the following expression: 

\vspace{-1mm}
\begin{equation}
    \small L_{G} = - E_{\Tilde{x} \sim P_{\Tilde{x}}}\big[ D(\Tilde{x},\phi) \big]
    \label{Loss G}
\end{equation}

\subsection{Architecture and training procedure}

To build our model we use convolutional neural networks (CNN) in one dimension. In the generator, we make use of nearest-neighbour upsampling layers to avoid artefacts. We set the kernel size $k = 5$, padding $p = 0$,  stride $s = 1$, with increasing dilation to enlarge the receptive field. Batch normalization is applied after each layer except for the penultimate one. Finally, $LeakyReLU(\cdot, \alpha = 0.2)$ activation function is employed, except in the output layer where we use a $Tanh(\cdot)$ activation, to constraint the output in the range $[-1,1]$.

%%%%----------------------------------------------------

The discriminator structure is composed of strided convolutions on which spectral normalization is used, as suggested in \cite{Gulrajani}. Dropout layers are triggered after each layer to regularise the discriminator, except for the first and last layers. The kernel size is set to 5  for all layers, with no padding and $LeakyReLU(\cdot, \alpha = 0.2)$ activation.

%%%%----------------------------------------------------

During the training, both the generator and the discriminator need to be updated at similar rates for stability and convergence. However, as the classification task is more challenging, we update the discriminator $5$ times per generator update. We employ RMSProp optimizer with a learning rate $ = 10^{-4}$ for both discriminator and generator, and we train it for $500$ epochs. Moreover, we employed $\lambda_1$= $5$, $\lambda_2$ = $5$, and dropout rate of $0.6$ for convergence.
More details about the architecture and the training can be found in \cite{Lopez:2022lkd}.

%%%%----------------------------------------------------

\section{Implemented features}

%%%%----------------------------------------------------

 \begin{figure}
     \centering
     \includegraphics[width=0.8\linewidth]{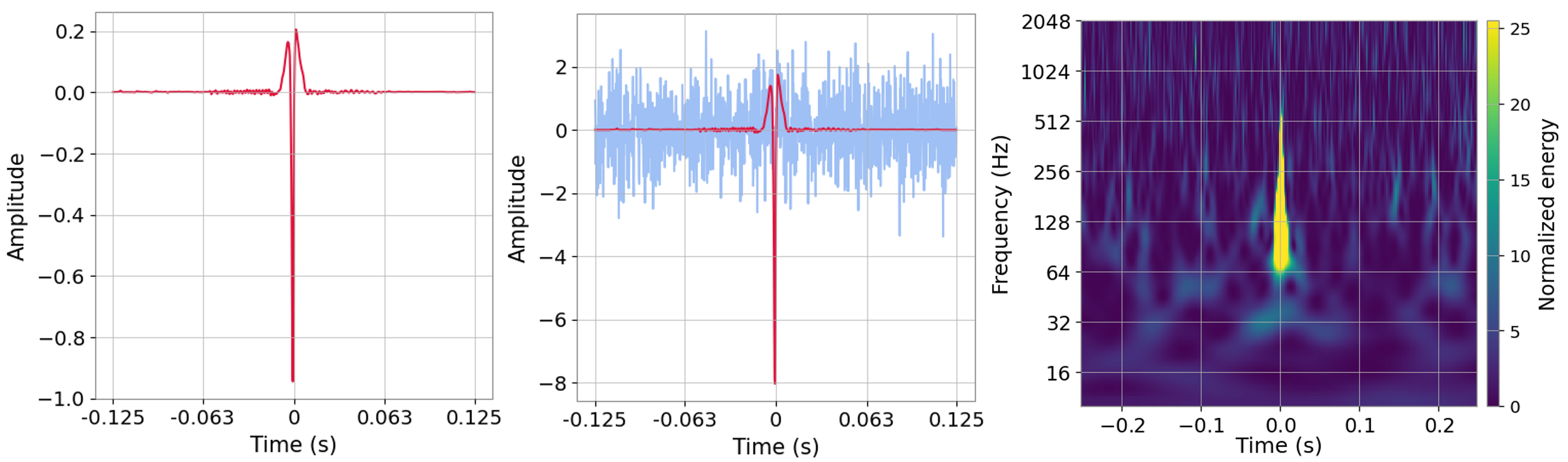}
     \caption{\textit{(Left)} Raw blip as a function of time. \textit{(Middle)} Raw blip injected in real (whitened) noise. \textit{(Right)}  Q-transform representation of injected glitch. Note the resemblance between the ``fake" glitch here and the real glitch in Fig.~\ref{fig:glitch_IMBH}.}
     \label{fig:example_glitch}
 \end{figure}

%%%%----------------------------------------------------

Once the model is trained, it is straightforward to use the generator network to produce random glitches. The output of the generator, a raw glitch, is the excess power of a \textit{whitened} time series evaluated on a fixed length time grid at a constant sampling rate $4096$ Hz and an amplitude in the range $[-1, 1]$. We can generate a raw glitch in $\sim 10 \text{ ms}$ on a laptop. 
A raw glitch can be straightforwardly injected into any whitened time series.
In fig.~\ref{fig:example_glitch} we plot an example of a raw glitch and of how it can be injected in white noise.

Since the amplitude of the generated glitches is arbitrary, before injecting the glitch into noise, a glitch has to be scaled to a user-given Signal-To-Noise (SNR) ratio $\rho$.
The SNR of a glitch $g(t)$ is defined as:

\vspace{-5mm}
\begin{equation}
    \rho^2 = 4 \int_{f_\text{min}}^{f_\text{max}} \text{d}f \frac{|\tilde{g}(f)|^2}{S_n(f)}
    \label{eq:snr}
\end{equation}
\vspace{-2mm}

%%%%----------------------------------------------------
\noindent
where again $\tilde{g}(f)$ is the glitch in the frequency domain and $S_n(f)$ is the actual PSD of the (whitened) data where the glitch is injected.
The raw glitch is then scaled to achieve the target SNR $\rho_{target}$ using: $g(t)^\prime = \frac{\rho_{target}}{\rho} g(t)$

The raw glitch can also be naturally \textit{resampled}\footnote[4]{When upsampling, we make the key assumption that there are no interesting features at frequencies higher than $4096$ Hz.} at the desired sampling rate. 
All such functionalities are implemented in the function \texttt{get\_raw\_glitch}: the interested user can find below an example of some working code.

%%%%----------------------------------------------------
\vspace{2mm}
\begin{lstlisting}
import gengli

g = gengli.glitch_generator('H1')
glitch = g.get_raw_glitch(snr = 10, srate = 2048.) 
\end{lstlisting}

%%%%----------------------------------------------------

%%%%----------------------------------------------------

\vspace{-3mm}
\subsection{{Selecting generations}}

 \begin{figure}
     \centering
     \includegraphics[width=0.5\linewidth]{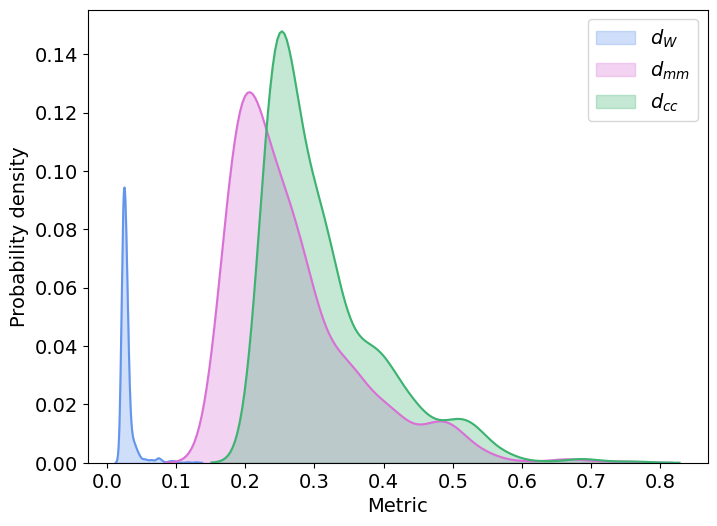}
     \caption{ Probability distribution of distances $(d_W, d_{mm}, d_{cc})$ for a benchmark population of $10^3$. }
     \label{fig:benchmark}
 \end{figure}
 
%%%%----------------------------------------------------
In \cite{Lopez:2022lkd}, it was shown that the generated glitches have the same statistical properties as the training population. However, the training data may (and most likely do) contain anomalous glitches that our model has also learnt to generate (see Section IV of \cite{Lopez:2022lkd}).
In this work, we propose a novel method to filter anomalous glitches.

To make this intuition mathematically precise, we start by considering three different measures of distances between an arbitrary pair of glitches:

%%%%----------------------------------------------------

\begin{itemize}
    \item Wasserstein distance $d_W$: a standard measure of distance between distributions, commonly employed in ML.
    \item Mis-match $d_{mm}$: a measure of distance based on the details of the filtering, standard in the GW field.
    \item Cross covariance $d_{cc}$:  we employ the quantity $1-k$, where $k$ is the normalized cross-covariance as defined in \cite{Lopez:2022lkd}.
\end{itemize}

%%%%----------------------------------------------------

We then populate a benchmark set of $N_b$ glitches with samples from the generator. For each of the $\frac{N_b(N_b-1)}{2}$ pairs of glitches in the benchmark set, we compute the three distances above. In Fig. \ref{fig:benchmark} we show the distribution for the three distances for a population of $N_b = 1000$.

%%%%----------------------------------------------------

For each new glitch being generated, we compute the set of average distances $(d_W, d_{mm}, d_{cc})$ between the glitch and the benchmark set, and we measure the set of percentiles $(p_W, p_{mm}, d_{cc})$ against the benchmark distances. The triplet $(p_W, p_{mm}, d_{cc})$ is our novelty measure for the glitch that will be used to filter glitches based on an anomaly score interval $[p_{min}, p_{max}]$. The code will output only glitches for which {\it all} the three anomaly scores lie within the interval.

%%%%----------------------------------------------------

\section{Conclusions}

%%%%----------------------------------------------------

We develop a methodology to generate artificial time-domain blip glitches data with the GAN algorithm. Because of the heavy pre-processing required to deal with real glitches (i.e. denoising, reconstructing etc...), this is a valid way to avoid using real data in state-of-the-art applications.
Due to the instability of GAN algorithms, in this particular research, we trained the model constructed in \cite{Gulrajani}, modified for time series, with the modified Wasserstein loss proposed in \cite{CTGAN}. This model has shown better performance in both training stability and accuracy.

%%%%----------------------------------------------------

The performance of the generated blips has been assessed in the companion paper \cite{Lopez:2022lkd}, where we employ several similarity distances: {Wasserstein distance ($d_{W}$), mismatch ($d_{mm}$) and cross-covariance ($d_{cc}$)}. The results of these metrics indicate that our model was able to learn the underlying distribution of blip glitches despite the presence of anomalies due to imperfections of the input data set. 

%%%%----------------------------------------------------

In this follow-up work, we introduce our open-source package \texttt{gengli}: it provides an easy-to-use interface to the trained GAN output, and has some additional features such as \textit{resampling} and/or \textit{scaling} a glitch and building a population with a given degree of ``anomaly".

Future work will condition GAN to other types of glitches (such as {\it koi-fish} and {\it tomte}), by using a wealth of data from O3.
Our work will enable the GW community to improve glitch classification with ML, study the properties of the glitch population and develop more realistic Mock Data challenges for studies on future GW detectors.

%%%%----------------------------------------------------

\section*{Acknowledgment}

%%%%----------------------------------------------------

 V.B. is supported by the Gravitational Wave Science (GWAS) grant funded by the French Community of Belgium, and M.L., S.C and S.S. are supported by the research program of the Netherlands Organisation for Scientific Research (NWO). Computational resources were provided by the LIGO Laboratory and supported by the National Science Foundation Grants No. PHY-0757058 and No. PHY-0823459. 

%%%%----------------------------------------------------

\end{document}